\documentclass[twocolumn,showpacs]{revtex4}

\usepackage{graphicx}%
\usepackage{dcolumn}
\usepackage{amsmath}

\begin{document}
\begin{sloppypar}

\title{Differential Cross Sections Measurement for the $pp\rightarrow
d\pi^+$ Reaction at 850 MeV/c}

\author{M. Betigeri$^{\ i}$,
J. Bojowald$^{\ a}$, A. Budzanowski$^{\ d}$, A. Chatterjee$^{\ i}$, J.
Ernst$^{\ g}$, L. Freindl$^{\ d}$, D. Frekers$^{\ h}$, W. Garske$^{\
h}$, K. Grewer$^{\ h}$, A. Hamacher$^{\ a}$, J. Ilieva$^{\ a,e}$, L.
Jarczyk$^{\ c}$, K. Kilian$^{\ a}$, S. Kliczewski$^{\ d}$, W.
Klimala$^{\ a,c}$, D. Kolev$^{\ f}$, T. Kutsarova$^{\ e}$, J. Lieb$^{\
j}$, H. Machner$^{\ a}$, A. Magiera$^{\ c}$, H. Nann$^{\ a}$, L.
Pentchev$^{\ e}$, D. Proti\'c$^{\ a}$, B. Razen$^{\ a}$, P. von
Rossen$^{\ a}$, B. J. Roy$^{\ i}$, R. Siudak$^{\ d}$, A.
Strza{\l}kowski$^{\ c}$, R. Tsenov$^{\ f}$, K. Zwoll$^{b}$}
\affiliation{
 a. Institut f\"{u}r Kernphysik, Forschungszentrum J\"{u}lich, J\"{u}lich, Germany
\\b. Zentrallabor f\"ur Elektronik, Forschungszentrum J\"ulich,
J\"ulich, Germany\\c. Institute of Physics, Jagellonian University,
Krakow, Poland\\d. Institute of Nuclear Physics, Krakow, Poland
\\e. Institute of Nuclear Physics and Nuclear Energy,  Sofia, Bulgaria
\\f. Physics Faculty, University of Sofia, Sofia, Bulgaria
\\g. Institut f\"ur Strahlen- und Kernphysik der Universit\"at
Bonn, Bonn, Germany
\\h. Institut f\"ur Kernphysik,  Universit\"at M\"unster,
M\"unster, Germany
\\i. Nuclear Physics Division, BARC, Bombay, India
\\j. Physics Department, George Mason University,
Fairfax, Virginia, USA} \collaboration{GEM Collaboration} \email{
h.machner@fz-juelich.de}

\date{\today }

\begin{abstract}
A stack of annular detectors made of high purity germanium and a magnet
spectrograph were used to measure $pp\rightarrow d\pi^+$ differential
cross sections at a beam momentum of 850 MeV/c over a large angular
range. A total cross section of $\sigma=0.2301\pm 0.0036\ (stat)\ \pm
0.0230\ (syst)$ mb and an anisotropy $A_2/A_0=0.856\pm0.016$ were
deduced. These values follow fits to low energy data. From the present
$A_2$ value it is found that the pionic p-wave amplitude $a_0$ is
larger than assumed so far.
\end{abstract}
\maketitle

\section[Introduction]{Introduction}

Although the reaction $pp\rightarrow d\pi^+$ has been intensively
studied, some open problems remain. These are mainly connected to the
importance of different partial wave amplitudes. For small pion
energies only three complex amplitudes are believed to contribute to
the physical observables: $a_0,\ a_1$ and $a_2$. Here the index denotes
the angular momentum in the proton--proton channel. Close to threshold,
the cross section is given by only one amplitude $a_1$ corresponding to
an s-wave in the final channel. For slightly larger energies, the
p-wave amplitudes $a_0$ and $a_2$ show up. Because $|a_2|$ is much
larger than $|a_0|$, the latter amplitude is often ignored \cite{Bugg}.
The d-wave amplitudes are even smaller. In the simplest experiments,
three observables can be measured, the total cross section, the
anisotropy and the asymmetry of the reaction products.  The usual
method to deduce the three complex amplitudes is then to make use of
the Watson theorem \cite{Wat54}. However, this procedure often yields
negative values for $|a_0|$ \cite{Hei99}. Because of its smallness, the
magnitude of $a_0$ cannot be extracted from the total cross section. It
enters the other two quantities via interference with other amplitudes.
Especially in the anisotropy, there is an interference between the two
p-waves. Thus the measurement of this quantity might yield an estimate
of $a_0$.

The energy dependence of the anisotropy, which will be defined later,
is a puzzle. Data published before 1996 like those of Ref.
\cite{Ritchie} lie on but often below the predictions of the SP96 phase
shift analysis carried out by the Virginia group \cite{SAID}. Near
threshold data, appeared first in 1996 \cite{Dro98,Hei96} and are in
agreement with the SP96 solution. The newest data are  from a
measurement of the inverse reaction \cite{Pasyuk}. They are even below
the Ritchie et al. data \cite{Ritchie}, thus indicating the negligible
effect of $a_0$ even in the interference term (see section
\ref{Results}). This is in contrast to SAID calculations as well as to
extrapolation of the new near threshold results. Furthermore, the new
data of Ref. \cite{Pasyuk} show a much larger scattering than the
threshold data, although the latter are from two different experiments.
It is worth mentioning that in the measurement of the anisotropy, the
luminosity cancels out. Large fluctuations in this quantity may thus
point to systematic errors other than target thickness and beam
intensity. We, therefore, investigate the situation by a measurement in
the range of these data not so close to threshold.

In the present experiment, the differential cross section for the
$pp\to d\pi^+$ reaction was measured at a beam momentum of 850 MeV/c,
which corresponds to a pion centre of mass momentum divided by the pion
rest mass of $\eta =0.51$. This value is within the range of the Pasyuk
et al. data \cite{Pasyuk} and will thus allow to proof the importance
of $pp$-wave interference in the anisotropy as already mentioned.

\section[Experimental Procedure]{Experimental Procedure}
The measurement of the reaction $pp\to d\pi^+$ was performed using a
detector that combines large momentum  and geometrical acceptance for
heavy recoiling reaction products. A proton beam with a momentum of 850
MeV/c was extracted from the COSY accelerator and focused onto a target
cell containing liquid hydrogen. It had a diameter of 6 mm and a
thickness of $6.4\pm 0.3$ mm \cite{Jae94} with windows of 1.5 $\mu$m
Mylar. The excellent ratio of hydrogen to heavier nuclei in the window
material reduced empty target events to a negligible level. The beam
spot had dimensions of less than 1 mm and divergences of better than 4
mrad. Beam halo events were suppressed using a plastic scintillator as
a veto counter with a 4­mm­diameter inner hole in front of the target.
The detector system is the ``Germanium Wall'', which is part of the
GEM-detector at COSY in J\"{u}lich \cite{Bet99} and the remodelled magnetic
spectrograph Big Karl \cite{Dro98}.  Here we give only some additional
details specific for this experiment.

The magnetic spectrograph was set to zero degrees in the laboratory
system. It was used to measure the low energy branch of deuterons close
to the primary beam. The germanium wall consists of three high purity
germanium detectors with radial symmetry with respect to the beam axis.
The first detector (called Quirl-detector) measures the position and
the energy loss of the penetrating particles. The active area of the
diode is divided on both sides into 200 segments by 200 grooves. Each
groove is shaped as an Archimedes' spiral covering an angular range of
$2 \pi$. They are mainly used for measuring the energy loss of the
penetrating particles or the total kinetic energy of stopped particles,
respectively. The next detectors are divided into 32 wedges to reduce
the counting rate per division leading to a higher maximum total
counting rate of the whole detector.

\begin{figure}[htbp]
\begin{center}
\includegraphics*[width= 9.cm]{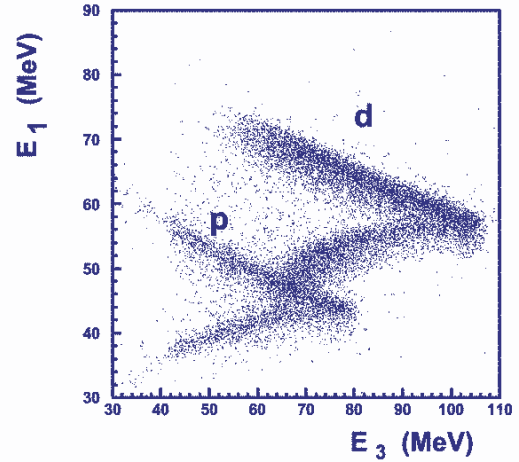}
\end{center}
\caption{\label{Delta_E} Energy loss in the first calorimeter crystal
$E_1$ versus the loss in the second named $E_3$ (A possible crystal
$E_2$ was not mounted in this experiment). The visible bands are due to
detected protons and deuterons.}
\end{figure}

Fig. \ref{Delta_E} shows the response of the germanium wall for
reaction particles. Clearly distinguished are two bands corresponding
to protons from the $pp\to pp\pi^0$ and deuterons from the $pp\to
d\pi^+$ reactions.
The quantities measured with the germanium wall are energy, emission
vertex and particle type. They were converted to a four-momentum
vector. These measurements and the knowledge of the four momenta in the
initial state yield the missing mass of the unobserved pion by applying
conservation of momentum, energy, charge and baryon number.

The germanium detectors have holes in their centres. The primary beam
passes through these holes and is then led via an exit in a side yoke
of the first dipole magnet of the magnetic spectrograph to a well
shielded beam dump. Recoiling deuterons at emission angles inside this
hole, i. e. close to zero degree, were detected by the magnetic
spectrograph. Details of such measurements are given in Ref.
\cite{Dro98}.

The reaction deuterons were selected in the off-line analysis by
applying gates to the kinematical loci in Fig. \ref{Delta_E}. The
efficiency of the analysis procedures were studied by Monte Carlo
calculations \cite{Gar99}. Finally, the data were corrected for reduced
efficiencies due to nuclear absorption in the detector material
\cite{Mac99}. The correction factors vary from 17\% to 22\% for the
present energy interval. The present setup has full acceptance for the
recoiling deuterons, except for a small area close to the beam exit
hole. For particles not stopped in the germanium wall, only energy loss
measurements are made. Because the angular distribution of the reaction
products must be symmetric to 90 degree in the centre of mass system,
it is sufficient to measure only one half of the distribution. The
deuterons emitted backward in the centre of mass system are all stopped
in the germanium wall. We restrict ourselves to this part of the
distribution in order to not introduce ambiguities when unfolding the
backbending part in Fig. \ref{Delta_E}.

Each of the measurements, i. e. emission angle $\Omega=(\theta,\phi)$
and energy of the recoiling deuteron, is sufficient to extract the
angular distribution. Because of this overconstraint, one can clearly
distinguish between reaction and background events.
\begin{figure}[htbp]
\begin{center}
\includegraphics*[width= 9.cm]{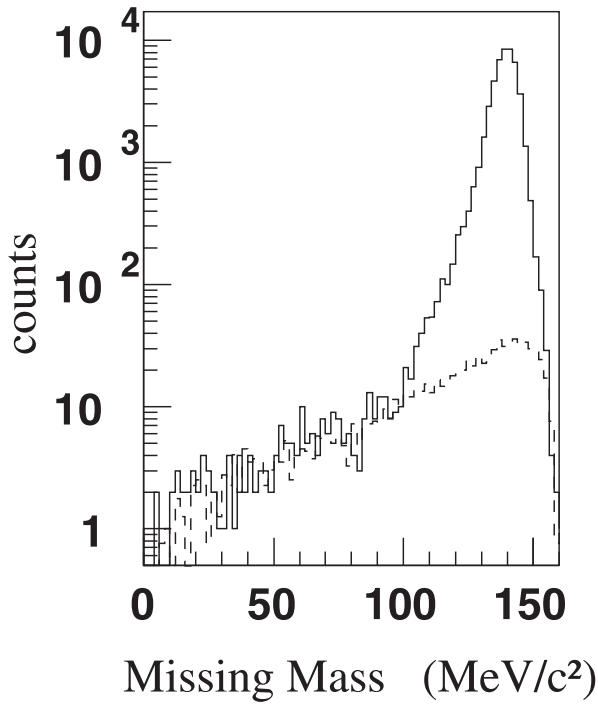}
\end{center}
\caption{\label{Missing_Mass}Missing mass spectrum. The experimental
result is shown by solid line histogram, the assumed background by
dashed line histogram.}
\end{figure}
This is shown in Fig. \ref{Missing_Mass}. The missing mass of the
unobserved $\pi^+$ is shown as deduced in the off line analysis. The
distribution has a resolution of $5.9\ MeV/c^2$. The logarithmic scale
was chosen to make the small background visible. The background, shown
by a dashed curve in Figure \ref{Missing_Mass}, was measured with an
empty target, normalized to integrated beam intensity and then
subtracted.

Finally, the counts were converted to cross sections by normalizing to
the target thickness and number of beam protons. The latter were
measured with calibrated luminosity monitor counters. In the
calibration procedure  the direct beam intensity and the corresponding
number of scattered particles are compared. The former was measured
with the trigger hodoscope in the focal plane of the magnet
spectrograph. This number is of course much larger and lead to dead
time in the hodoscope. The beam intensity was then reduced by
debunching the beam between the ion source and the cyclotron injector.
For sufficiently small beam intensity the relation between monitors and
hodoscope is linear. The counting rate in the monitors was in the the
production runs small enough to have dead time effects on a negligible
level. This procedure yields a systematic uncertainty of $5\%$ for the
beam intensity. This yields together with the target thickness
uncertainty of also $5\%$ a maximal total systematical error of $10\%$
for the overall normalization of the differential cross section when
both errors are added linearly.

\section [Results]{Results}\label{Results}
\begin{figure}[htbp]
 \begin{center}
\includegraphics*[width= 7.cm]{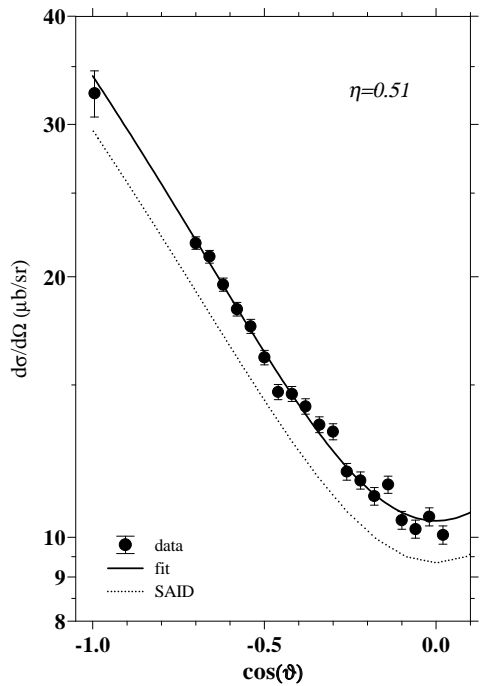}
 \end{center}
\caption{\label{ang_dist} The measured angular distribution is shown by
the full symbols. Shown are the statistical errors only. The overall
normalization of the cross section has total systematical error of
$10\%$. The solid curve is a Legendre polynomial fit up to second
order, the dotted curve the SP96 solution \cite{SAID}.}
\end{figure}

The measured angular distribution in the centre of mass system are
shown in Fig. \ref{ang_dist}.
The point near $\cos (\theta )=-1$ was measured with the magnetic
spectrograph. It has larger statistical error than all other points
which were measured with the germanium wall. This is due to its small
acceptance. The data show a very strong anisotropy when compared to
those close to the threshold \cite{Dro98,Hei96}. This is an indication
of the p-wave strength and thus of the importance of the intermediate
$\Delta$-p system. The angular distribution was fitted by a series of
Legendre polynomials $P_{2L}[\cos(\theta)]$
\begin{equation}\label{Legendre}
4\pi \frac{{d\sigma (\theta )}}{{d\Omega }} = A_0  + A_2 P_2 [\cos
(\theta )] + A_4 P_4 [\cos (\theta )].
\end{equation}
The fitted parameters $A_{2L}$ are connected with amplitudes $a_i$
\begin{equation}\label{A_0}
A_0 = \sigma = {\textstyle{1 \over 4}}[|a_0 |^2  + |a_1 |^2  + |a_2 |^2
+ C(|d|^2 )]
\end{equation}
where $C(|d|^2)$ represents all d-wave contributions,
\begin{eqnarray}\label{A_2}
A_2  =&& \frac{1}{4}\left| {a_2 } \right|^2  + \frac{3}{{14}}\left|
{a_6 } \right|^2  - \sqrt {{\raise0.5ex\hbox{$\scriptstyle 1$}
\kern-0.1em/\kern-0.15em \lower0.25ex\hbox{$\scriptstyle 2$}}} \Re
\left( {a_0 a_2^* } \right)\nonumber\\&& + \sqrt
{{\raise0.5ex\hbox{$\scriptstyle 1$} \kern-0.1em/\kern-0.15em
\lower0.25ex\hbox{$\scriptstyle 8$}}} \Re \left( {a_1 a_3^* } \right) +
\sqrt {{\raise0.5ex\hbox{$\scriptstyle 5$} \kern-0.1em/\kern-0.15em
\lower0.25ex\hbox{$\scriptstyle 8$}}} \Re \left( {a_1 a_4^* }
\right)\nonumber\\&& + \frac{1}{2}\sqrt
{{\raise0.5ex\hbox{$\scriptstyle 5$} \kern-0.1em/\kern-0.15em
\lower0.25ex\hbox{$\scriptstyle 7$}}} \Re \left( {a_1 a_5^* } \right) +
\sqrt {{\raise0.5ex\hbox{$\scriptstyle 1$} \kern-0.1em/\kern-0.15em
\lower0.25ex\hbox{$\scriptstyle 2$}}} \Re \left( {a_1 a_6^* }
\right)\nonumber\\&& + D\left( {\left| d \right|^2 } \right) + E\left(
{dd} \right)
\end{eqnarray}
with $D(|d|^2)$ denoting other d-wave contributions and $E(dd)$ terms
with interferences between two d-waves. $A_4$ is given by
\begin{eqnarray}\label{A_4}
A_4  = && - \frac{5}{{49}}|a_5 |^2  + \frac{1}{{28}}|a_6 |^2  +
\frac{9}{{14}}\Re (a_3 a_6^* )\nonumber\\&& + \frac{{10}}{{49}}\sqrt
{14} \,\Re (a_4 a_5^* ) + \frac{5}{{14}}\sqrt 5 \,\Re (a_4 a_6^*
)\nonumber\\&& + \frac{5}{7}\sqrt {\frac{5}{{14}}} \,\Re (a_5 a_6^* ).
\end{eqnarray}

The amplitude $a_1$ is
the s-wave in the pion-deuteron channel, $a_0$ and  $a_2$ denote
p-waves and  $a_3$ to  $a_6$ denote d-waves. In Eqs. \ref{A_0} to
\ref{A_4} the notation of Mandl and Regge \cite{Man55} is used. The
total cross section can near threshold be written as
\begin{equation}\label{Centrifugal}
\sigma = \alpha_0 \eta+\alpha_1 \eta^3.
\end{equation}
The first term corresponds to the s-wave amplitude $a_1$ and the second
to the two p-wave amplitudes $a_0$ and $a_2$.

Fits of Eq. \ref{Legendre} up to second and to fourth order were
performed. Although the $\chi^2$-values of both fits are similar, the
large uncertainties in $A_4$ and in $A_2$ favour the second-order fit.
Application of F-statistics indicates the same conclusion. The results
are independent whether the point at $\cos (\theta )=-1$ is excluded or
not. The negative sign of $A_4$ is in agreement with a theoretical
prediction \cite{Niskanen} as well as with phase-shift analyses
\cite{Bugg,SAID}. On the other hand, corresponding fits to data in this
energy range \cite{Ritchie} yield positive values. From phase-shift
analysis it is known that all terms in d-waves are small. Only
$\Re(a_6)$ is of some size. If we neglect all other terms one gets
\begin{eqnarray}
A_4  = \frac{1}{{28}}\left\{ {|a_6 |^2  + \Re \left[ {18(a_3 a_6^* ) +
10\sqrt 5 (a_4 a_6^* )} \right]} \right\}.
\end{eqnarray}
The first term is positive and the second negative. This yields a
cancelation making the resulting value even smaller in agreement with
the present result. For the sake of simplicity we rely on the second
order fit. This choice may introduce a systematical error. However, the
change in the quantity of interest, i. e. the ratio $A_2/A_0$ is small;
both results agree with each other within error bars. The second order
fit is also shown in Fig. \ref{ang_dist} as solid curve.
\begin{table}[h]
\caption{\label{Leg_fits} Fitted Legendre polynomial coefficients
(Eq.
 \ref{Legendre}) in $\mu$b}
\begin{center}
\begin{tabular}{|c|c|c|}
 \hline
  parameter & second order  & fourth order   \\ \hline
  $A_0$ & $230.1\pm1.2$ &  $228.5\pm1.5$  \\
  $A_2$ & $197.0\pm3.1$ &  $190.8\pm5.1$  \\
  $A_4$ &  & $-6.02\pm3.9$ \\ \hline
\end{tabular}
\end{center}
\end{table}
Since $A_0$ is the total cross section we have from the second order
fit a value $\sigma=230.1\pm 3.6\,(stat) \, \pm 23.0\,(sys)\ \mu b$
with the systematical uncertainty discussed above. The efficiency
correction is angle dependent, varying from 17\% to 22\% in the
presently measured range. If a 10\% precision of this correction is
assumed, it adds 1.5\% to the uncertainty in the total cross section
only. This uncertainty is added in quadrature to the fit error and is
contained in the statistical uncertainty. We compare the present
angular distribution with a prediction from the SAID phase shift
analysis \cite{SAID}, shown as dotted curve in Fig. \ref{ang_dist}. The
prediction is always $\approx 10\%$ below the data. This is just the
sum of all systematic errors of the present experiment. Whether this is
the reason for the disagreement or not will be discussed below. From
the result in the Table an anisotropy $A_2/A_0=0.856\pm0.014\,
(fit)\,\pm0.0045\,(effic.\,corr.)$ is obtained. This is close to the
SAID prediction of $0.89$ indicating the good reproduction of the shape
of differential cross sections.

\begin{figure}[htbp]
 \begin{center}
\includegraphics*[width= 10.cm]{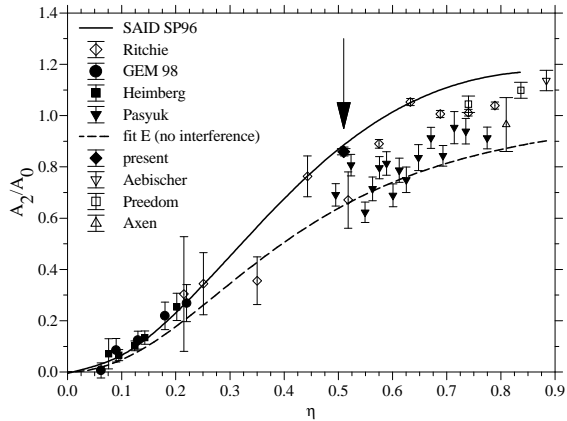}
 \end{center}
\caption{\label{anisotropy}Deduced ratio $A_2/A_0$ as function of the
dimensionless pion momentum.  The presently deduced value is indicated
by the arrow. The earlier data are indicated by different symbols. The
SAID calculation is shown as solid curve, the calculation with the
assumption of no interference as dashed curve.}
\end{figure}

This ratio is compared in Fig. \ref{anisotropy} with previous results
\cite{Aebischer,Axen,Predom}. The point joins the data in the upper
part of the band formed by the experiments and is in agreement with the
extrapolation of the low energy results assuming a dependence
$A_2\propto \eta ^3$. However, it is $15\%$ larger than the Pasyuk et
al. result. This seems to be a large discrepancy since only the shape
of the angular distributions are compared. Also shown in Fig.
\ref{anisotropy} is the prediction from fit E in Ref. \cite{Mac98}
under the assumption of no interference at all: $A_2/A_0=\alpha_1
\eta^3/(\alpha_0 \eta + \alpha_1 \eta^3)$ (compare Eq.
\ref{Centrifugal}).

It should be mentioned that the quality of fits for the total cross
sections for the phase shift analysis and fit E from Ref.
\cite{Mac98}is similar, except for the near threshold region. However,
the corresponding data are in the fit E but not in the SAID analysis.
We have also inspected the results of the phase shift analysis solution
C500. It only differs very slightly in the range $\eta <0.35$ from the
solution SP96  and gives almost identical results for larger
$\eta$-values. It is worth mentioning that in Eq. \ref{A_2}, according
to SAID,  the term $\Re(a_0a_2^*)$ is several orders of magnitude
larger than all the others, except the $|a_2|^2$ term.

\section{Discussion}
An almost complete angular distribution for the reaction $pp\to d\pi^+$
at a beam momentum of 850 MeV/c was measured. This corresponds to a
pion centre of mass momentum $\eta=p_\pi/m_\pi=0.51$. The measurement
was performed with a solid state detector with axial symmetry for
larger angles and a magnetic spectrograph for emission angles close to
the primary beam. The angular distribution shows a large anisotropy
when compared to data closer to the threshold. This is an indication
that p-wave emission is the dominant reaction mechanism. Because d-wave
amplitudes are much smaller than the s- and p-wave amplitudes one
cannot extract their strength from the total cross section. The angular
distribution is more sensitive to these waves. The present data give
almost no evidence for a d-wave contribution to the angular
distribution. Precise measurements at somewhat higher energies are
desirable to investigate the importance of d-waves esp. the sign of the
Legendre coefficient $A_4$ in the range below the resonance. The ratio
$A_2/A_0$ does not demand d-waves. However, they contribute to the
scattering asymmetry via interferences with p-waves having larger
amplitudes. The weakness of d-wave amplitudes was also recently found
in the measurement of spin transfer coefficients even at 400 MeV beam
energy \cite{Prz99}.

The present measurement confirms a larger ratio $A_2/A_0$ as extracted
from recent measurements at lower beam momenta and is in good agreement
with the phase shift analysis of the Virginia group. It disagrees with
the results of Ref. \cite{Pasyuk}. However, these data show a large
scatter which is surprising, since the comparison is made on a relative
base. It has its origin mainly in large fluctuations of the $A_2$
values. The agreement between the new data and the SP96 solution, which
does not include these data, is much better than the earlier phase
shift analysis \cite{Bugg}. The present total cross section
$\sigma=A_0$ and the second Legendre coefficient $A_2$ are larger than
the SP96 solution but the ratio $A_2/A_0$ is in excellent agreement
with this solution. A non negligible value for $a_0$, as is applied in
the SP96 solution, is confirmed by the present measurement. One needs
more observables than the presently measured ones to deduce the
amplitudes. Since the present result is in agreement with the SP96
solution we will rely on this solution. It yields for the ratio
$|a_0|/|a_2|$ a value of 0.099 close to threshold and 0.096 for the
present beam momentum. This can be compared to the result obtained by
Ref. \cite{Hei96}. From cross section and polarisation measurements,
which, together with the above cited Watson theorem, a ratio
$|a_0|/|a_2|=(9\pm 36)*10^{-3}$ was extracted , a value which is
compatible with zero.

\section{Acknowledgement}
We are grateful to the COSY operation crew for their efforts making a
good beam. Support by BMBF Germany (06 MS 568 I TP4), Internationales
B\"{u}ro des BMBF (X081.24 and 211.6), SCSR Poland (2P302 025 and 2P03B 88
08), and COSY J\"{u}lich is gratefully acknowledged.

\end{sloppypar}
\end{document}